\documentclass[12pt,epsf]{article}

\usepackage{times}
\usepackage{graphicx}
\usepackage{amsfonts}
\usepackage{amsmath}
\usepackage{amssymb}
\usepackage{amstext}
\usepackage{amsthm}

\usepackage{epsfig}
\usepackage{color}
\psfigdriver{dvips}
\usepackage{latexsym}
\usepackage[matrix,curve]{xypic}
\usepackage{rotating}
\rotdriver{dvips}

\begin{document}

\baselineskip 6mm
\renewcommand{\thefootnote}{\fnsymbol{footnote}}


\newcommand{\nc}{\newcommand}
\newcommand{\rnc}{\renewcommand}


\rnc{\baselinestretch}{1.24}    
\setlength{\jot}{6pt}       
\rnc{\arraystretch}{1.24}   

\makeatletter
\rnc{\theequation}{\thesection.\arabic{equation}}
\@addtoreset{equation}{section}
\makeatother



\nc{\be}{\begin{equation}}

\nc{\ee}{\end{equation}}

\nc{\bea}{\begin{eqnarray}}

\nc{\eea}{\end{eqnarray}}

\nc{\xx}{\nonumber\\}

\nc{\ct}{\cite}

\nc{\la}{\label}

\nc{\eq}[1]{(\ref{#1})}

\nc{\newcaption}[1]{\centerline{\parbox{6in}{\caption{#1}}}}

\nc{\fig}[3]{

\begin{figure}
\centerline{\epsfxsize=#1\epsfbox{#2.eps}}
\newcaption{#3. \label{#2}}
\end{figure}
}


\def\CA{{\cal A}}
\def\CC{{\cal C}}
\def\CD{{\cal D}}
\def\CE{{\cal E}}
\def\CF{{\cal F}}
\def\CG{{\cal G}}
\def\CH{{\cal H}}
\def\CK{{\cal K}}
\def\CL{{\cal L}}
\def\CM{{\cal M}}
\def\CN{{\cal N}}
\def\CO{{\cal O}}
\def\CP{{\cal P}}
\def\CR{{\cal R}}
\def\CS{{\cal S}}
\def\CU{{\cal U}}
\def\CV{{\cal V}}
\def\CW{{\cal W}}
\def\CY{{\cal Y}}
\def\CZ{{\cal Z}}


\def\IB{{\hbox{{\rm I}\kern-.2em\hbox{\rm B}}}}
\def\IC{\,\,{\hbox{{\rm I}\kern-.50em\hbox{\bf C}}}}
\def\ID{{\hbox{{\rm I}\kern-.2em\hbox{\rm D}}}}
\def\IF{{\hbox{{\rm I}\kern-.2em\hbox{\rm F}}}}
\def\IH{{\hbox{{\rm I}\kern-.2em\hbox{\rm H}}}}
\def\IN{{\hbox{{\rm I}\kern-.2em\hbox{\rm N}}}}
\def\IP{{\hbox{{\rm I}\kern-.2em\hbox{\rm P}}}}
\def\IR{{\hbox{{\rm I}\kern-.2em\hbox{\rm R}}}}
\def\IZ{{\hbox{{\rm Z}\kern-.4em\hbox{\rm Z}}}}


\def\a{\alpha}
\def\b{\beta}
\def\d{\delta}
\def\ep{\epsilon}
\def\ga{\gamma}
\def\k{\kappa}
\def\l{\lambda}
\def\s{\sigma}
\def\t{\theta}
\def\w{\omega}
\def\G{\Gamma}


\def\half{\frac{1}{2}}
\def\dint#1#2{\int\limits_{#1}^{#2}}
\def\goto{\rightarrow}
\def\para{\parallel}
\def\brac#1{\langle #1 \rangle}
\def\curl{\nabla\times}
\def\div{\nabla\cdot}
\def\p{\partial}


\def\Tr{{\rm Tr}\,}
\def\det{{\rm det}}


\def\vare{\varepsilon}
\def\zbar{\bar{z}}
\def\wbar{\bar{w}}
\def\what#1{\widehat{#1}}


\def\ad{\dot{a}}
\def\bd{\dot{b}}
\def\cd{\dot{c}}
\def\dd{\dot{d}}
\def\so{SO(4)}
\def\bfr{{\bf R}}
\def\bfc{{\bf C}}
\def\bfz{{\bf Z}}

\begin{titlepage}


\hfill\parbox{3.7cm} {{\tt arXiv:1412.1757}}

\vspace{15mm}

\begin{center}
{\Large \bf Mirror Symmetry in Emergent Gravity}

\vspace{10mm}

Hyun Seok Yang \footnote{hsyang@kias.re.kr} \footnote{Present address: Center for Quantum Spacetime, 
Sogang University, Seoul 04107, Korea}
\\[10mm]

{\sl School of Physics, Korea Institute for Advanced Study,
Seoul 130-722, Korea}

\end{center}

\thispagestyle{empty}

\vskip1cm


\centerline{\bf ABSTRACT}
\vskip 4mm
\noindent

Given a six-dimensional symplectic manifold $(M, B)$, a nondegenerate, co-closed four-form $C$
introduces a dual symplectic structure $\widetilde{B} = *C $ independent of $B$ via the Hodge duality $*$.
We show that the doubling of symplectic structures due to the Hodge duality results in
two independent classes of noncommutative $U(1)$ gauge fields by considering the Seiberg-Witten map
for each symplectic structure. As a result, emergent gravity suggests a beautiful picture that
the variety of six-dimensional manifolds emergent from noncommutative $U(1)$ gauge fields is doubled.
In particular, the doubling for the variety of emergent Calabi-Yau manifolds allows us to arrange
a pair of Calabi-Yau manifolds such that they are mirror to each other.
Therefore, we argue that the mirror symmetry of Calabi-Yau manifolds is the Hodge theory
for the deformation of symplectic and dual symplectic structures.
\\



\vspace{1cm}

\today

\end{titlepage}

\renewcommand{\thefootnote}{\arabic{footnote}}
\setcounter{footnote}{0}

\section{Introduction}

Emergent gravity is based on a novel form of the equivalence principle known as the
Darboux theorem or the Moser lemma in symplectic geometry stating that the electromagnetic
force can always be eliminated by a local coordinate transformation as far as
spacetime admits a symplectic structure, in other words, a microscopic spacetime
becomes noncommutative (NC) \cite{hsy-ijmp09,hsy-jhep09,q-emg}.
See also closely related works \cite{rivelles,berens,h-stein} and reviews \cite{emg-review,hsy-review}.
A basic idea of emergent gravity is to isomorphically map the deformations of symplectic structure
on a symplectic manifold $(M, B)$ to those of Riemannian metric on an emergent Riemannian
manifold $(\mathcal{M}, \mathcal{G})$. The deformation of symplectic structure is described by
considering a line bundle $L \to M$ over the symplectic manifold $(M, B)$ such that the curvature
$F=dA$ of the line bundle $L$ defines a locally deformed symplectic two-form $\mathcal{F} = B + F$
on an open neighborhood $\mathcal{U} \subset M$. The resulting symplectic
structure $(M, \mathcal{F})$ defines a dynamical system and it can be quantized using the underlying
Poisson structure $\theta \equiv B^{-1}$. The quantization of the dynamical system results in
a dynamical NC spacetime which is described by a NC $U(1)$ gauge theory and interpreted as
the quantization of the emergent Riemannian manifold $(\mathcal{M}, \mathcal{G})$ \cite{q-emg}.

The dynamical variables described by the NC $U(1)$ gauge theory form a NC algebra $\mathcal{A}_\theta$
under a quantum $\star$-bracket $[-,-]$. Given a quantum algebra $(\mathcal{A}_\theta, [-,-])$,
there are two isomorphic representations of the quantum algebra. Since the NC algebra $\mathcal{A}_\theta$
admits a separable Hilbert space $\mathcal{H}$ and so a countable basis, 
there is a Lie algebra homomorphism $\rho: \mathcal{A}_\theta \to \mathrm{End}(\mathcal{H})$ which 
is a matrix representation in the Hilbert space $\mathcal{H}$.
An element in $\mathrm{End}(\mathcal{H})$ is an $N \times N$ matrix where $N=\mathrm{dim}(\mathcal{H})$ and,
in our case, $N \to \infty$. Another representation is given by the adjoint representation $\mathfrak{ad}:
\mathcal{A}_\theta \to \mathfrak{D}$ which is also a Lie algebra homomorphism.
Since $\mathfrak{D}$ is the inner derivation of the algebra $\mathcal{A}_\theta$
under the $\star$-bracket $[-,-]$, an element in $\mathfrak{D}$ is given by a differential operator
in the differential algebra. An interesting problem is to identify the theories described
by the set of large $N$ matrices in $\mathrm{End}(\mathcal{H})$ and the set of differential operators
in $\mathfrak{D}$, respectively. It turns out \cite{hsy-jhep09,q-emg} that the former is described by
a large $N$ gauge theory in lower dimensions whose dynamical variables take values in $\mathrm{End}(\mathcal{H})$
and the latter in a classical limit describes a higher-dimensional gravity.
Since these two theories are equally derived from the NC $U(1)$ gauge theory taking values
in the NC algebra $\mathcal{A}_\theta$, they should be physically equivalent (or dual)
to each other. The relationship between a lower-dimensional large $N$ gauge theory and
a higher-dimensional gravity is known as the gauge/gravity duality or large $N$ duality.

Emergent gravity is very similar to mirror symmetry in the sense that the deformation of symplectic
structure is isomorphically mapped to the deformation of Riemannian metric \cite{hsy-jhep09}.
In particular, a K\"ahler manifold carries a natural symplectic structure inherited
from a K\"ahler metric. Therefore the K\"ahler manifold is a symplectic manifold.
Mirror symmetry in string theory is a correspondence between two topologically distinct Calabi-Yau (CY)
manifolds $(X, Y)$ that give rise to the exactly same physical theory \cite{mirror-book}.
The idea is that mirror symmetry provides an isomorphism between certain aspects of complex geometry
on $X$ and certain aspects of symplectic geometry on $Y$.
In particular, the homological mirror symmetry \cite{kontsevich} states that the derived category of coherent
sheaves on a K\"ahler manifold should be isomorphic to the Fukaya category of a mirror symplectic manifold.
The Fukaya category is described by the Lagrangian submanifolds of a given symplectic manifold as its objects
and the Floer homology groups as their morphisms. Since emergent gravity is aimed at constructing
a Riemannian geometry from a symplectic geometry, it will be interesting to see how the mirror symmetry
in string theory is realized from the emergent gravity approach.

Recently we showed \cite{hea,cy-hsy} that a holomorphic line bundle with a nondegenerate
curvature two-form of rank $2n$ is equivalent to a $2n$-dimensional K\"ahler manifold.
This relation was first found and explored in the context of topological strings in \cite{ivno}.
In particular, CY $n$-folds for $n$ = 2 and 3 are emergent from the commutative limit of
NC $U(1)$ instantons in four and six dimensions, respectively, where quantum algebra reduces to
a classical Poisson algebra. Since CY manifolds are derived from the commutative limit of
NC Hermitian $U(1)$ instantons, it should be interesting to investigate how to realize the mirror symmetry
of CY manifolds from the NC $U(1)$ gauge theory according to the emergent gravity picture.
The basic idea is to observe that there are two independent sources of symplectic structures in six dimensions
as was briefly explained in Ref. \cite{cy-hsy}.
Since a symplectic manifold $(M, B)$ is an orientable manifold, so admits a globally defined volume form,
one can introduce the Hodge dual operator $*: \Omega^k (M) \to \Omega^{6-k} (M)$ between $k$-form and
$(6-k)$-form vector spaces. In particular, this implies that the vector space of two-forms $\Lambda^2 M =
\Omega^2 (M) \oplus * \Omega^{4} (M)$ is doubled since one can get a two-form $\widetilde{B} = * C
\in \Omega^{2} (M)$ by taking the Hodge-dual of a four-form $C \in \Omega^{4} (M)$.
It can be shown that the new two-form $\widetilde{B} = * C$ is a symplectic two-form if the four-form $C$
is co-closed and nondegenerate. The symplectic structure $(M, \widetilde{B})$ is completely independent of
the original one $(M, B)$. Since emergent geometry is derived from the deformation of symplectic structures
and this deformation can be represented by NC $U(1)$ gauge fields via the Darboux theorem,
the symplectic structure $(M, \widetilde{B})$ will also generate (dual) NC $U(1)$ gauge fields and
give rise to a corresponding (dual) emergent geometry, as illustrated in Eqs. \eq{dualdbi-id1} and \eq{dualdbi-id2}.
In this paper we will explore the physical consequence of the doubling of sympectic structures 
from the emergent geometry perspective.

In this paper, we explore the relation between the NC $U(1)$ gauge theory in six dimensions and
the K\"ahler gravity on a non-compact CY threefold and identify the curvature of a holomorphic line bundle
with the K\"ahler form for a CY manifold. We observe that, due to the nontrivial four-form on a CY manifold, 
one can define the second dual holomorphic line bundle whose curvature is related to the Hodge-dual of 
the four-form and argue that the CY manifold emergent from the dual holomorphic line bundle is mirror 
to the CY manifold emergent from the ordinary holomorphic line bundle.
In fact, this relation has been found and studied in \cite{ivno} and 
the similar question has been further discussed in the papers \cite{noteadd1,noteadd2,noteadd3},
although the language is a bit different, involving the counting of cohomologies via topological strings
or the duality transformations of D-branes of different codimensions.
It was also noted in \cite{noteadd2,noteadd3} that the duality in \cite{ivno} follows from the S-duality of
type IIB superstrings. Although the duality transformation implies essentially the same result as ours
for the mirror relation in CY manifolds, it has not been surfaced yet to explicitly show how to construct 
the mirror pair in terms of $U(1)$ gauge theory. Our paper fills out the gap by the explicit gauge theory 
construction of the mirror pair and our result is obtained without using the machinery of 
topological string theory and the conjectured web of dualities in string theory.
Since the duality found in \cite{ivno} can be reformulated from the emergent gravity perspective,
it is hoped that our gauge theory realization of the mirror pairs sheds light on the duality 
between Gromov-Witten invariants of a CY threefold $X$ and a topologically
twisted NC $U(1)$ gauge theory on $X$.

This paper is organized as follows. In section 2, we recapitulate the basic idea on the
K\"ahler manifolds emergent from a holomorphic line bundle with a nondegenerate curvature
two-form \cite{hea,cy-hsy}. The emergent CY manifolds are derived in a more elegant way by realizing
the emergent gravity using the associative algebra of NC $U(1)$ gauge fields.
In section 3, we observe that the variety of six-dimensional manifolds emergent from NC
$U(1)$ gauge fields is doubled thanks to the isomorphism between two-forms and four-forms by the
Hodge duality. Therefore the emergent gravity in six dimensions suggests a beautiful picture that
the doubling for the variety of emergent CY manifolds allows
us to arrange a pair of CY manifolds such that they are mirror to each other.
In section 4, we argue that two CY manifolds arising from a line bundle and a dual line bundle
are a mirror pair and thus the mirror symmetry of CY manifolds is
the Hodge theory for the deformation of symplectic and dual symplectic structures.
In section 5, we discuss some generalization of emergent K\"ahler manifolds.

\section{Holomorphic Line Bundle and K\"ahler Manifolds}

Let $\pi: L \to M$ be a line bundle over a six-dimensional complex manifold $M$ whose connection
is denoted by $\mathcal{A} = \mathcal{A}_a (x) dx^a$. The curvature of the line bundle $L$ is
given by $\mathcal{F} = \frac{1}{2} \mathcal{F}_{ab} dx^a \wedge dx^b = d \mathcal{A}$.
Given a complex structure of the base manifold $M$, one can canonically decompose the curvature
two-form as
\begin{equation}\label{cdec-f}
    \mathcal{F} = \mathcal{F}^{(2,0)} \oplus \mathcal{F}^{(1,1)} \oplus \mathcal{F}^{(0,2)}.
\end{equation}
A holomorphic line bundle $L$ over $M$ is a complex vector bundle of (complex) rank one admitting
holomorphic transition functions \cite{grha-pag}. A line bundle $L$ becomes a holomorphic
line bundle if
\begin{equation}\label{holo-l}
\mathcal{F}^{(2,0)} = \mathcal{F}^{(0,2)} = 0.
\end{equation}
Accordingly, the curvature of a holomorphic line bundle $L$ consists of a $(1,1)$-form only, i.e.,
$\mathcal{F} = \mathcal{F}^{(1,1)}$.

For simplicity, we will assume that $M = \mathbb{C}^3$, whose complex coordinates are given by
\begin{equation}\label{comp-coord}
    z^i = y^{2i-1} + \sqrt{-1} y^{2i}, \qquad \overline{z}^{\bar{i}} = y^{2i-1} - \sqrt{-1} y^{2i}, \qquad
    i, \bar{i} = 1, 2, 3.
\end{equation}
Later we will briefly discuss a generalization to a compact complex manifold, e.g.,
$M = T^6 \cong T_{\mathbb{C}}^3$, a three-dimensional complex torus.
According to the complex coordinates in Eq. \eq{comp-coord}, the connection of $L$,
which is called $U(1)$ gauge fields, also takes the decomposition given by
\begin{equation}\label{comp-u1g}
    \mathcal{A}_i = \frac{1}{2} \bigl( \mathcal{A}_{2i-1} - \sqrt{-1} \mathcal{A}_{2i} \bigr),
    \qquad \mathcal{A}_{\overline{i}}
    = \frac{1}{2} \bigl( \mathcal{A}_{2i-1} + \sqrt{-1} \mathcal{A}_{2i} \bigr).
\end{equation}
Then the field strengths of $(2,0)$ and $(1,1)$ parts in Eq. \eq{cdec-f} are, respectively,
given by
\begin{eqnarray} \label{fs-20}
&& \mathcal{F}_{ij} = \frac{1}{4} \bigl( \mathcal{F}_{2i-1, 2j-1} -  \mathcal{F}_{2i, 2j} \bigr)
- \frac{\sqrt{-1}}{4} \bigl( \mathcal{F}_{2i-1, 2j} +  \mathcal{F}_{2i, 2j-1} \bigr), \\
\label{fs-11}
&& \mathcal{F}_{i\overline{j}} = \frac{1}{4} \bigl( \mathcal{F}_{2i-1, 2j-1} +  \mathcal{F}_{2i, 2j} \bigr)
+ \frac{\sqrt{-1}}{4} \bigl( \mathcal{F}_{2i-1, 2j} -  \mathcal{F}_{2i, 2j-1} \bigr).
\end{eqnarray}
Therefore, the curvature of a holomorphic line bundle, i.e., $\mathcal{F}_{ij} =
\mathcal{F}_{\overline{i}\overline{j}} = 0$, must obey the relation
\begin{equation}\label{holo-lcomp}
\mathcal{F}_{2i-1, 2j-1} =  \mathcal{F}_{2i, 2j}, \qquad
\mathcal{F}_{2i-1, 2j} = - \mathcal{F}_{2i, 2j-1}, \qquad i, j= 1,2,3.
\end{equation}
Since $\mathcal{F}_{ij} = \partial_i \mathcal{A}_j - \partial_j \mathcal{A}_i$ and
$\mathcal{F}_{\bar{i}\bar{j}} = \overline{\partial}_{\bar{i}} \mathcal{A}_{\overline{j}}
- \overline{\partial}_{\bar{j}} \mathcal{A}_{\overline{i}}$, the condition \eq{holo-l}
for the holomorphic line bundle can be solved by
\begin{equation}\label{holo-gauge}
\mathcal{A}_i = - \frac{\sqrt{-1}}{2} \partial_i \phi(z, \overline{z}), \qquad
\mathcal{A}_{\overline{i}} = \frac{\sqrt{-1}}{2} \overline{\partial}_{\bar{i}} \phi(z, \overline{z})
\end{equation}
where $\phi(z, \overline{z})$ is a real smooth function on $\mathbb{C}^3$.
Then the field strength of a holomorphic line bundle is given by
\begin{equation}\label{holo-field}
 \mathcal{F}_{i\overline{j}} = \sqrt{-1} \partial_i \overline{\partial}_{\bar{j}}  \phi(z, \overline{z}).
\end{equation}

Suppose that $\mathcal{M}$ is a six-dimensional complex manifold with the metric $ds^2 = \mathcal{G}_{\mu\nu} (x)
dx^\mu dx^\nu$. On a complex manifold, the metric can also be decomposed into three types:
\begin{equation}\label{c-metric}
\mathcal{G}_{\mu\nu} = \mathcal{G}_{\alpha\beta} \oplus \mathcal{G}_{\alpha\overline{\beta}}
\oplus \mathcal{G}_{\overline{\alpha}\overline{\beta}},
\end{equation}
where we have split a curved space index $\mu = 1, \cdots, 6 = (\alpha, \bar{\alpha})$
into a holomorphic index $\alpha = 1, 2, 3$ and an anti-holomorphic one $\bar{\alpha} = 1, 2, 3$,
similarly a tangent space index $a = 1, \cdots, 6 = (i, \bar{i})$ into $i = 1, 2, 3$
and $\bar{i} = 1, 2, 3$. A complex manifold $\mathcal{M}$ is called a Hermitian
manifold \cite{besse} if
\begin{equation}\label{herm-manf}
\mathcal{G}_{\alpha\beta} = \mathcal{G}_{\overline{\alpha}\overline{\beta}} = 0.
\end{equation}
In terms of real components, the Hermitian condition \eq{herm-manf} means that
\begin{equation}\label{herm-comp}
\mathcal{G}_{2\alpha-1, 2\beta-1} =  \mathcal{G}_{2\alpha, 2\beta}, \qquad
\mathcal{G}_{2\alpha-1, 2\beta} = - \mathcal{G}_{2\alpha, 2\beta-1}, \qquad \alpha, \beta = 1,2,3,
\end{equation}
which looks similar to Eq. \eq{holo-lcomp} although one is antisymmetric and the other is symmetric.
After all, the Hermitian metric consists of $(1,1)$-type only, i.e.,
\begin{equation}\label{herm-metric}
    ds^2 = \mathcal{G}_{\alpha\overline{\beta}} (z, \overline{z}) dz^\alpha d\overline{z}^{\bar{\beta}}.
\end{equation}
Given a Hermitian metric, one can introduce a fundamental two-form defined by
\begin{equation}\label{fund-form}
    \Omega = \sqrt{-1} \mathcal{G}_{\alpha\overline{\beta}} (z, \overline{z}) dz^\alpha
    \wedge d\overline{z}^{\bar{\beta}}.
\end{equation}
A K\"ahler manifold is defined as a Hermitian manifold with a closed fundamental two-form,
i.e., $d \Omega = 0$ \cite{besse}. The so-called K\"ahler condition, $d \Omega = 0$,
can be solved by the metric given by
\begin{equation}\label{k-metric}
 \mathcal{G}_{\alpha\overline{\beta}} = \partial_{\alpha} \overline{\partial}_{\bar{\beta}} K(z, \overline{z}),
\end{equation}
where $K(z, \overline{z})$ is a real smooth function on a complex manifold $\mathcal{M}$ and
is called a K\"ahler potential.

Now let us look at the curvature $\mathcal{F}$ of a holomorphic line bundle and
the K\"ahler form $\Omega$ of a K\"ahler manifold that are, respectively, given by
\begin{eqnarray} \label{curv-holo}
&&  \mathcal{F} = \sqrt{-1} \partial_i \overline{\partial}_{\bar{j}}  \phi(z, \overline{z})
 dz^i \wedge d\overline{z}^{\bar{j}} = \sqrt{-1} \partial \overline{\partial} \phi(z, \overline{z}), \\
 \label{om-kah}
&& \Omega = \sqrt{-1} \partial_{\alpha} \overline{\partial}_{\bar{\beta}} K(z, \overline{z})
dz^\alpha \wedge d\overline{z}^{\bar{\beta}}= \sqrt{-1} \partial \overline{\partial} K(z, \overline{z}).
\end{eqnarray}
Since $ \phi(z, \overline{z})$ and $K(z, \overline{z})$ are arbitrary smooth functions
on a complex manifold in addition to a striking superficial similarity of $\mathcal{F}$ and $\Omega$,
an innocent question naturally arises whether it is possible to identify them or when we can identity them.
If one recalls that the K\"ahler form $\Omega$ is a symplectic structure, then the answer may be obvious.
The curvature two-form $\mathcal{F}$ must be a symplectic structure to make sense the identification.
Indeed, it was shown in \cite{hea,ivno} that one can identify $\phi(z, \overline{z})$ with $K(z, \overline{z})$
if the curvature $\mathcal{F}$ of a holomorphic line bundle is a symplectic structure, i.e.,
a nondegenerate, closed two-form. A nondegenerate two-form $\mathcal{F} = \frac{1}{2}
\mathcal{F}_{ab} (x) dx^a \wedge dx^b$ means that $\det \mathcal{F}_{ab} (x) \neq 0, \; \forall x \in M$.
Of course, it is not a typical situation in the Maxwell's electromagnetism where
$\mathcal{F}_{ab} (x)|_{|x| \to \infty} \to 0$. To emphasize the nondegenerateness of the field strength,
let us represent it by
\begin{equation}\label{dec-ndfield}
   \mathcal{F} = B + F
\end{equation}
where $B \equiv \mathcal{F} |_{|x| \to \infty}$ is a nowhere vanishing two-form of rank 6 and $F = dA$.
The identification of $\phi(z, \overline{z})$ with $K(z, \overline{z})$ means that
a holomorphic line bundle with a nondegenerate curvature two-form of rank 6 is equivalent to
a six-dimensional K\"ahler manifold. Then the real function $\phi(z, \overline{z})$ and so $K(z, \overline{z})$
will be determined by solving the equations of motion of $U(1)$ gauge fields.
In other words, (generalized) Maxwell's equations for $U(1)$ gauge fields on a holomorphic line bundle
can be translated into Einstein's equations for a K\"ahler manifold.
For example, one may wonder what is the gauge theory object that gives rise to a CY manifold
which is a K\"ahler manifold with a vanishing first Chern class.
It was verified in \cite{hea,cy-hsy} that CY $n$-folds for $n=2$ and $3$ are emergent
from the commutative limit of NC $U(1)$ instantons in four and six dimensions, respectively.

Let us recapitulate why it is possible to make the identification up to holomorphic
gauge transformations:\footnote{\label{gluing}Note that both $\phi(z, \overline{z})$ and $K(z, \overline{z})$
are locally defined. They may not fit together on the overlap $\mathcal{U}_i \cap \mathcal{U}_j$ to
give a globally defined function on a complex manifold $M$ where
$\bigcup_i (\mathcal{U}_i, z^{(i)})$ is a holomorphic atlas of $M$ \cite{grha-pag}.
However, the curvature $\mathcal{F}$ and the K\"ahler form $\Omega$ can be globally defined.
For example, one can use a $U(1)$ gauge transformation, $\mathcal{A} \to \mathcal{A} + d f$
where $f \in C^\infty (M)$, to glue locally defined functions $\phi^{(i)}$ on each coordinate
patch $\mathcal{U}_i$. On the overlap $\mathcal{U}_i \cap \mathcal{U}_j$ of two coordinate patches,
the $U(1)$ gauge transformation reads as $\phi^{(i)} = \phi^{(j)} + f^{(ij)} (z)
+ \overline{f^{(ij)}} (\overline{z})$ where two real functions $\phi^{(i)}$ and $\phi^{(j)}$
are defined on $\mathcal{U}_i$ and $\mathcal{U}_j$, respectively \cite{jsw-ncl}.
This gluing of $U(1)$ gauge fields can be translated into that of K\"ahler potentials
according to the identification \eq{iden=fk}.}
\begin{equation}\label{iden=fk}
 \phi(z, \overline{z}) = K(z, \overline{z}),
\end{equation}
if the curvature $\mathcal{F}$ of a holomorphic line bundle is regarded as a symplectic structure on $M$.
An important fact is that a symplectic structure, for instance, the $B$-field in Eq. \eq{dec-ndfield},
provides a bundle isomorphism $B: TM \to T^* M$ by $X \mapsto A = \iota_X B$ where
$X \in \Gamma(TM)$ is an arbitrary vector field, since $B$ is a nondegenerate two-form.
As a result, the field strength in Eq. \eq{dec-ndfield} can be written as
\begin{equation}\label{btr-f}
\mathcal{F} = (1+ \mathcal{L}_X) B \approx e^{\mathcal{L}_X} B,
\end{equation}
where $\mathcal{L}_X = d \iota_X + \iota_X d$ is the Lie derivative with respect to the vector field $X$.
Since a vector field is an infinitesimal generator of local coordinate transformations, in other words,
a Lie algebra generator of diffeomorphisms $\mathrm{Diff}(M)$, the result \eq{btr-f}
implies \cite{hsy-ijmp09,hsy-jhep09} that it is possible to find a local coordinate
transformation $\phi \in \mathrm{Diff}(M)$ eliminating
dynamical $U(1)$ gauge fields in $\mathcal{F}$ such that $\phi^* (\mathcal{F}) = B$, i.e.,
$\phi^* = (1+ \mathcal{L}_X)^{-1} \approx e^{-\mathcal{L}_X}$.
This statement is known as the Darboux theorem or the Moser lemma in symplectic geometry \cite{sg-book}.
It is arguably a novel form of the equivalence principle for the electromagnetic force.
This fact leads to a remarkable conclusion \cite{q-emg} that, in the presence of $B$-fields,
the ``dynamical" symplectic manifold $(M, \mathcal{F})$ respects a (dynamical) diffeomorphism symmetry
generated by the vector field $X \in \Gamma(TM)$, so the underlying local gauge symmetry is rather enhanced.
Here we mean the ``dynamical" for fluctuating fields around a background like Eq. \eq{dec-ndfield}.
Therefore, we fall into a situation similar to general relativity that the dynamical symplectic
manifold $(M, \mathcal{F})$ can be locally trivialized by
a coordinate transformation $\phi \in \mathrm{Diff}(M)$.

In terms of local coordinates, the coordinate transformation $\phi \in \mathrm{Diff}(M)$ may
be represented by
\begin{equation}\label{phi}
    \phi: y^a \mapsto x^a (y) = y^a + \theta^{ab} a_b (y)
\end{equation}
where $\theta \equiv B^{-1}$ and the dynamical coordinates $a_b (y)$ will be called symplectic
gauge fields. By using the above coordinates, the Darboux transformation
obeying $\phi^* (\mathcal{F}) = B$ is explicitly written as
\begin{equation}\label{darboux}
    \big( B_{ab} + F_{ab}(x) \big) \frac{\partial x^a}{\partial y^\mu} \frac{\partial x^b}{\partial y^\nu}
    = B_{\mu\nu},
\end{equation}
where $B$ is assumed to be constant without loss of generality.
Since both sides of Eq. \eq{darboux} are invertible, one can deduce \cite{cornalba,jur-sch,liu} that
\begin{eqnarray}\label{sw-darboux}
    \Theta^{ab}(x) \equiv (\mathcal{F}^{-1})^{ab} (x) &=& \{ x^a (y), x^b (y) \}_\theta \nonumber \\
    &=& \big( \theta (B - f) \theta \big)^{ab} (y),
\end{eqnarray}
where we have introduced the Poisson bracket defined by
\begin{equation}\label{poisson-br}
    \{ \psi (y), \varphi (y) \}_\theta = \theta^{\mu\nu}
    \frac{\partial \psi (y)}{\partial y^\mu} \frac{\partial \varphi (y)}{\partial y^\nu}
\end{equation}
for $\psi, \varphi \in C^\infty (M)$ and the field strength of symplectic gauge fields is given by
\begin{equation}\label{field-symg}
    f_{ab} (y) = \partial_a a_b (y) - \partial_b a_a (y) + \{a_a (y), a_b (y) \}_\theta.
\end{equation}

The identification \eq{iden=fk} suggests a fascinating path for the quantization of K\"ahler manifolds.
Note that the symplectic manifold $(M, \mathcal{F})$ is a dynamical system since it can be understood
as the deformation of a symplectic manifold $(M, B)$ by the electromagnetic force $F = dA$.
Thus one may quantize the dynamical system of the symplectic manifold $(M, \mathcal{F})$ rather than
trying to quantize a K\"ahler manifold directly \cite{q-emg}.\footnote{\label{foot-ivno}We have come
to a notice that the basic idea on the emergent K\"ahler manifold in this paper is essentially the same as
the realization of K\"ahler gravity in terms of $U(1)$ gauge theory presented in a beautiful paper \cite{ivno}.
The authors in \cite{ivno} conclude that for topological strings the $U(1)$ gauge theory is the fundamental description of gravity at all scales including the Planck scale, where it leads to a quantum gravitational foam.}
The quantization $\mathcal{Q}$ is straightforward as the dynamical system equips with
the intrinsic Poisson structure \eq{poisson-br} like as quantum mechanics.
An underlying math is essentially the same as quantum mechanics.
It results in a NC $U(1)$ gauge theory \cite{ncft-sw} on a quantized or NC space,
denoted by $\mathbb{R}^{6}_\theta$, whose coordinate generators satisfy the commutation relation
\begin{equation}\label{nc-space}
    [y^a, y^b] = i \theta^{ab}.
\end{equation}
The NC $\star$-algebra generated by the Moyal-Heisenberg algebra (\ref{nc-space}) will be denoted
by $\mathcal{A}_\theta$ \cite{ncft-rev}. The NC $U(1)$ gauge theory is constructed by lifting
the coordinate transformation (\ref{phi}) to a local automorphism of $\mathcal{A}_\theta$ defined
by $\mathcal{Q}: \phi \mapsto \mathcal{D}_A$ which acts on the NC coordinates $y^a$ as \cite{jur-sch,juscwe}
\begin{equation}\label{dyna-codi}
    \mathcal{D}_A (y^a) \equiv  \widehat{X}^a (y) = y^a + \theta^{ab} \widehat{A}_b (y)
 \in \mathcal{A}_\theta.
\end{equation}
It ascertains that NC $U(1)$ gauge fields are obtained by quantizing symplectic gauge fields,
i.e., $\widehat{A}_a = \mathcal{Q} (a_a)$. Upon quantization, the Poisson bracket is similarly
lifted to a NC bracket in $\mathcal{A}_\theta$.
For example, the Poisson bracket relation (\ref{sw-darboux}) is now defined by the commutation relation
\begin{equation}\label{dnc-space}
    [\widehat{X}^a, \widehat{X}^b ]_\star = i \big( \theta(B - \widehat{F}) \theta \big)^{ab},
\end{equation}
where the field strength of NC $U(1)$ gauge fields $\widehat{A}_a$ is given by
\begin{equation}\label{nc-curvature}
    \widehat{F}_{ab} = \partial_a \widehat{A}_b - \partial_b \widehat{A}_a
    -i [\widehat{A}_a, \widehat{A}_b]_\star.
\end{equation}
Here we observe \cite{q-emg} that NC $U(1)$ gauge fields describe a dynamical NC
spacetime (\ref{dnc-space}) which is a deformation of the vacuum NC spacetime (\ref{nc-space}).
To sum up, a dynamical NC spacetime is defined by the quantization of a line bundle $L$ over
a symplectic manifold $(M, B)$ and described by a NC $U(1)$ gauge theory.

The identification \eq{iden=fk} attains its vitality in the following way.
It can be shown \cite{hea} that the Darboux transformation \eq{darboux} leads to a remarkable
identity between Dirac-Born-Infeld densities (up to total derivatives):
\begin{eqnarray}\label{dbi-id1}
\sqrt{\det(g + \mathcal{F})}  &=& \sqrt{\det(\mathcal{G} + B)}  \\
\label{dbi-id2}
&=& \frac{g_s}{G_s} \sqrt{\det(G + \widehat{F} - B)},
\end{eqnarray}
where the flat metrics $(g, G)$ are the K\"ahler metric of $\mathbb{C}^3$ and
$B$ its K\"ahler form. The identity \eq{dbi-id1} clearly verifies that $U(1)$ gauge fields
on the left-hand side must be a connection of a holomorphic line bundle obeying
$\mathcal{F}_{ij} = \mathcal{F}_{\overline{i}\overline{j}} = 0$ to give rise to
a K\"ahler metric $\mathcal{G}$ on the right-hand side and vice versa. Thus it establishes
the identification \eq{iden=fk}.\footnote{\label{i-sheaf}Since $U(1)$ instantons
on a commutative space $M$ are singular and this singularity is resolved
in the NC description of $U(1)$ instantons \cite{nc-inst,hsyang-nci},
it is necessary to allow singular $U(1)$ gauge fields on $M$
to make sense the identification \eq{iden=fk} in a general context.
To admit such a singular gauge field, we need to relax the notion of the line bundle $L$.
The natural replacement for the holomorphic line bundle $L$ is the rank one torsion free
sheaf with the same first Chern class \cite{ivno}. We will assume the generalization of
the line bundle by allowing singular $U(1)$ gauge fields.} This demonstration is also true
for the identity \eq{dbi-id2}; a connection on a NC holomorphic line bundle obeying $\widehat{F}^{(2,0)}
= \widehat{F}^{(0,2)} = 0$ gives rise to a K\"ahler metric $\mathcal{G}$ \cite{hea,cy-hsy}.

Although the identities in Eqs. \eq{dbi-id1} and \eq{dbi-id2} clearly illustrate how to realize
the gauge/gravity duality using a NC $U(1)$ gauge theory based on an associative
algebra $\mathcal{A}_\theta$, the $\star$-algebra $\mathcal{A}_\theta$ provides a more elegant approach
for the gauge/gravity duality. A preliminary step to derive gravitational variables from NC $U(1)$ gauge
fields \cite{hsy-jhep09,q-emg,hsy-review} is to note that
the NC $\star$-algebra $\mathcal{A}_\theta$ admits a nontrivial inner automorphism $\mathfrak{A}$
defined by $\mathcal{O} \mapsto \mathcal{O}' = U \star \mathcal{O} \star U^{-1}$
where $U \in \mathfrak{A}$ and $\mathcal{O} \in \mathcal{A}_\theta$. Its infinitesimal generators
consist of an inner derivation $\mathfrak{D}$. Then the inner derivation $\mathfrak{D}$ manifests
a well-known Lie algebra homomorphism defined by the map
\begin{equation}\label{der-map}
 \mathcal{A}_\theta \to \mathfrak{D}:
 \mathcal{O} \mapsto \mathrm{ad}_{\mathcal{O}} = -i [\mathcal{O}, \cdot \; ]_\star
\end{equation}
for any $\mathcal{O} \in \mathcal{A}_\theta$. Using the Jacobi identity of the NC $\star$-algebra
$\mathcal{A}_\theta$, it is easy to verify the Lie algebra homomorphism:
\begin{equation}\label{lie-homo}
 [ \mathrm{ad}_{\mathcal{O}_1}, \mathrm{ad}_{\mathcal{O}_2} ] =
 -i \mathrm{ad}_{[\mathcal{O}_1, \mathcal{O}_2]_\star}
\end{equation}
for any $\mathcal{O}_1, \mathcal{O}_2 \in \mathcal{A}_\theta$.
In particular, we define the set of NC vector fields given by
\begin{equation}\label{nc-vector}
\{ \widehat{V}_a \equiv \mathrm{ad}_{\widehat{D}_a} \in \mathfrak{D}| \widehat{D}_a (y)
 = p_a + \widehat{A}_a (y) \in \mathcal{A}_\theta, \; a = 1, \cdots, 6 \}
\end{equation}
where $p_a = B_{ab} y^b$ and $\widehat{D}_a (y) \equiv  \mathcal{D}_A (p_a) = B_{ab} \widehat{X}^b (y)$.
One can apply the Lie algebra homomorphism \eq{lie-homo} to
the commutation relation
\begin{equation}\label{ncc-dd}
    -i [\widehat{D}_a, \widehat{D}_b ]_\star = -B_{ab} + \widehat{F}_{ab}
\end{equation}
to yield the relation \cite{hsy-jhep09,hsy-review,hsy-jpcs12}
\begin{equation}\label{homo-f}
\mathrm{ad}_{\widehat{F}_{ab}}  =  [\widehat{V}_a, \widehat{V}_b] \in \mathfrak{D}.
\end{equation}

The identification \eq{iden=fk} may be confirmed by using the Lie algebra of derivations
in Eqs. \eq{nc-vector} and (\ref{homo-f}). To be precise, the derivation $\mathfrak{D}$ of the associative
algebra $\mathcal{A}_\theta$ defined by a NC $U(1)$ gauge theory is associated with a (quantized) frame
bundle of an emergent spacetime manifold $\mathcal{M}$ \cite{q-emg}. For example, we recently
verified a particular case of the identity \eq{iden=fk} \cite{hea,cy-hsy} that the commutative limit of
six-dimensional NC Hermitian $U(1)$ instantons obeying the Hermitian Yang-Mills
equation \cite{non-inst,bkp-hbps,hsy-epjc09}
\begin{equation}\label{nchym1}
    \widehat{F}  = - * (\widehat{F} \wedge B),
\end{equation}
where $B = \frac{1}{2} I_{\mu\nu} dy^\mu \wedge dy^\nu$ is the K\"ahler form of $\mathbb{C}^3$,
is equivalent to CY manifolds obeying the (local) Einstein equation,
$\det  \mathcal{G}_{\alpha\overline{\beta}}=1$.

\section{Doubling of Emergent Calabi-Yau Manifolds}

We observed in the previous section that emergent gravity is defined by considering the deformation of
a symplectic manifold $(M, B)$ by a line bundle $L \to M$. The line bundle $L$ results in a dynamical
symplectic manifold $(M, \mathcal{F})$ by introducing a new symplectic structure $\mathcal{F} = B + F$
where $F = dA$ is identified with the curvature of the line bundle \cite{q-emg}.
It is important to note \cite{sg-book} that a symplectic manifold $(M, B)$ is necessarily an orientable
manifold since the symplectic structure $B$ admits a nowhere vanishing volume form $\nu = \frac{1}{3!} B^3$.
Then a globally defined volume form introduces the Hodge-dual operation $*: \Omega^k (M) \to \Omega^{6-k} (M)$
between vector spaces of $k$-forms and $(6-k)$-forms. This implies that the vector space $\Lambda^2 M$
of two-forms is enlarged twice:
\begin{equation}\label{double-2form}
 \Lambda^2 M = \Omega^2 (M) \oplus * \Omega^{4} (M),
\end{equation}
since there are additional two-forms from the Hodge-dual of four-forms in $\Omega^{4} (M)$
in addition to the original two-forms in $\Omega^{2} (M)$.
Let $C$ be a nondegenerate four-form that is co-closed,
i.e., $\delta C = 0$ where
\begin{equation}\label{adjoint-d}
\delta = - * d *: \Omega^k (M) \to \Omega^{k-1} (M)
\end{equation}
is the adjoint exterior differential operator \cite{grha-pag,besse}. Define a two-form
$\widetilde{B} \equiv * C$. Then we have the relation
\begin{equation}\label{co-close}
    \delta C = 0 \quad \Leftrightarrow \quad d\widetilde{B} = 0.
\end{equation}
Therefore $\widetilde{B}$ defines another symplectic structure independent of $B$.
Hence it should be possible to consider the deformation of the {\it dual} symplectic
structure $\widetilde{B}$ by incorporating a dual line bundle $\widetilde{L} \to M$.
Then an interesting question is what is a physical consequence of the doubling of symplectic structures
in Eq. \eq{double-2form} due to the Hodge duality.

Let $\widetilde{A}$ be a $U(1)$ connection of the dual line bundle $\widetilde{L}$ and
$\widetilde{F} = d \widetilde{A}$ its curvature. According to the vector space structure
in Eq. \eq{double-2form}, we identify the curvature $\widetilde{F} = d \widetilde{A}$ with
the Hodge-dual of a four-form $G$, i.e., $\widetilde{F} = *G$. The Bianchi identity for
$\widetilde{L}$ is then equal to the co-closedness of the four-form $G$:
\begin{equation}\label{co-bianchi}
    d\widetilde{F} = 0 \quad \Leftrightarrow \quad \delta G = 0.
\end{equation}
Using the nilpotency $\delta^2 = 0$ \cite{besse}, the so-called co-Bianchi identity, $\delta G = 0$,
can locally be solved by $G = \delta D$ and the connection $\widetilde{A}$ of
the dual line bundle $\widetilde{L}$ can be identified with the Hodge-dual of the five-form connection $D$,
viz., $\widetilde{A} = - * D$. As the usual line bundle $L$ over a symplectic manifold $(M, B)$,
the dual line bundle $\widetilde{L}$ will similarly deform the dual symplectic structure $\widetilde{B}$
of the base manifold $M$, leading to a new symplectic structure
\begin{equation}\label{dual-symp}
\widetilde{\mathcal{F}} \equiv \widetilde{B} + \widetilde{F}= *(C+G).
\end{equation}
Hence the dual line bundle $\widetilde{L}$ also results in
a dynamical symplectic manifold $(M, \widetilde{\mathcal{F}})$. Recall that the symplectic gauge fields
in Eq. \eq{phi} have been introduced via a Darboux transformation $\phi \in \mathrm{Diff} (M)$ such that
$\phi^* (\mathcal{F}) = B$. Similarly, one can consider a local coordinate transformation $\widetilde{\phi}
\in \mathrm{Diff}(M)$ such that $\widetilde{\phi}^* (\widetilde{\mathcal{F}}) = \widetilde{B}$.
Let us introduce Darboux coordinates $u^a \; (a = 1, \cdots, 6)$ so that the coordinate transformation $\widetilde{\phi} \in \mathrm{Diff}(M)$ is given by
\begin{equation}\label{phi-tilde}
    \widetilde{\phi}: u^a \mapsto w^a (u) = u^a + \widetilde{\theta}^{ab} c_b (u)
\end{equation}
where $\widetilde{\theta} \equiv \widetilde{B}^{-1}$ and the dynamical coordinates $c_b (u)$
will be called {\it dual} symplectic gauge fields. By using the Darboux coordinates, the coordinate
transformation obeying $\widetilde{\phi}^* (\widetilde{\mathcal{F}}) = \widetilde{B}$ is explicitly
written as
\begin{equation}\label{dual-darboux}
    \big( \widetilde{B}_{ab} + \widetilde{F}_{ab}(w) \big) \frac{\partial w^a}{\partial u^\mu}
    \frac{\partial w^b}{\partial u^\nu}
    = \widetilde{B}_{\mu\nu}.
\end{equation}
It should be emphasized that the dual symplectic gauge fields in Eq. \eq{phi-tilde} are completely
independent of the symplectic gauge fields in Eq. \eq{phi} to be compatible with
the doubling of the vector space in Eq. \eq{double-2form}.

The dual Poisson structure $\widetilde{\theta} = \widetilde{B}^{-1}$ defines a new Poisson bracket
given by
\begin{equation}\label{dual-poibr}
    \{ \psi (u), \varphi (u) \}_{\widetilde{\theta}} = \widetilde{\theta}^{\mu\nu}
    \frac{\partial \psi (u)}{\partial u^\mu} \frac{\partial \varphi (u)}{\partial u^\nu}
\end{equation}
for $\psi, \varphi \in C^\infty (M)$. From the Darboux transformation \eq{dual-darboux},
one can then deduce the Poisson bracket relation
\begin{eqnarray}\label{dual-swd}
    (\widetilde{\mathcal{F}}^{-1})^{ab} (w) &=& \{ w^a (u), w^b (u) \}_{\widetilde{\theta}} \nonumber \\
    &=& \big( \widetilde{\theta} (\widetilde{B} - \widetilde{f}) \widetilde{\theta} \big)^{ab} (u),
\end{eqnarray}
where the field strength of dual symplectic gauge fields is defined by
\begin{equation}\label{field-symd}
    \widetilde{f}_{ab} (u) = \partial_a c_b (u) - \partial_b c_a (u)
    + \{ c_a (u), c_b (u) \}_{\widetilde{\theta}}
\end{equation}
with $\partial_a := \frac{\partial}{\partial u^a}$.
The quantization $\mathcal{Q}$ of the dynamical symplectic manifold $(M, \widetilde{\mathcal{F}})$
is defined by canonically quantizing a Poisson algebra $(C^\infty (M), \{ -, - \}_{\widetilde{\theta}})$ \cite{q-emg}. It leads to another NC $\star$-algebra $\mathcal{A}_{\widetilde{\theta}}$ generated by
the Moyal-Heisenberg algebra satisfying the commutation relation
\begin{equation}\label{dual-ncspace}
    [u^a, u^b] = i \widetilde{\theta}^{ab}.
\end{equation}
The NC $\star$-algebra $\mathcal{A}_{\widetilde{\theta}}$ is independent of the previous one
$\mathcal{A}_{\theta}$ by our construction.
For example, a local automorphism of $\mathcal{A}_{\widetilde{\theta}}$ defined by $\mathcal{Q}:
\widetilde{\phi} \mapsto \mathcal{D}_{\widetilde{A}}$ acts on
the NC coordinates $u^a$ as \cite{jur-sch,juscwe}
\begin{equation}\label{dy-dualcod}
    \mathcal{D}_{\widetilde{A}} (u^a) \equiv  \widehat{W}^a (u) = u^a
    + \widetilde{\theta}^{ab} \widehat{C}_b (u) \in \mathcal{A}_{\widetilde{\theta}},
\end{equation}
where $\widehat{C}_a = \mathcal{Q}(c_a)$ are another NC $U(1)$ gauge fields obtained by
quantizing dual symplectic gauge fields $c_a(u)$. The covariant dynamical coordinates $\widehat{W}^a (u)$
satisfy the commutation relation
\begin{equation}\label{nc-dspace}
    [\widehat{W}^a, \widehat{W}^b ]_{\widetilde{\star}} = i \big( \widetilde{\theta}(\widetilde{B}
    - \widehat{H}) \widetilde{\theta} \big)^{ab},
\end{equation}
where the field strength of NC $U(1)$ gauge fields $\widehat{C}_a$ is given by
\begin{equation}\label{nc-dcurv}
    \widehat{H}_{ab} = \partial_a \widehat{C}_b - \partial_b \widehat{C}_a
    -i [\widehat{C}_a, \widehat{C}_b]_{\widetilde{\star}}.
\end{equation}

In consequence, there exist two independent NC $\star$-algebras to define a dynamical NC spacetime.
They are separately obtained by quantizing the line bundles $L$ and $\widetilde{L}$ describing
the deformation of symplectic structures in $\Omega^2 (M)$ and $* \Omega^4 (M)$, respectively.
Since the two vector spaces in Eq. \eq{double-2form} are isomorphic to each other so that
they should be treated on an equal footing, the exactly same argument for the previous
symplectic manifold $(M, \mathcal{F})$ can be equally applied to the dual symplectic
manifold $(M, \widetilde{\mathcal{F}})$. It is straightforward to derive from
the Darboux transformation \eq{dual-darboux} the following identity
between Dirac-Born-Infeld densities (up to total derivatives) \cite{hea}:
\begin{eqnarray}\label{dualdbi-id1}
\sqrt{\det(\widetilde{g} + \widetilde{\mathcal{F}})}  &=&
\sqrt{\det(\widetilde{\mathcal{G}} + \widetilde{B})}  \\
\label{dualdbi-id2}
&=& \frac{\widetilde{g}_s}{\widetilde{G}_s} \sqrt{\det(\widetilde{G} + \widehat{H} -\widetilde{B})},
\end{eqnarray}
where $(\widetilde{g}, \widetilde{G})$ are the K\"ahler metric of $\mathbb{C}^3$ and
$\widetilde{B}$ its K\"ahler form and $(\widetilde{g}_s, \widetilde{G}_s)$ are coupling constants
in the dual gauge theories. The identity \eq{dualdbi-id1} immediately verifies that $U(1)$ gauge fields
on the left-hand side must be a connection of a holomorphic line bundle obeying
$\widetilde{\mathcal{F}}_{ij} = \widetilde{\mathcal{F}}_{\overline{i}\overline{j}} = 0$
to give rise to a K\"ahler metric $\widetilde{\mathcal{G}}_{\alpha\overline{\beta}} = \partial_{\alpha} \overline{\partial}_{\bar{\beta}} \widetilde{K}(z, \overline{z})$ on the right-hand side where
$\widetilde{K}(z, \overline{z})$ is the K\"ahler potential of a K\"ahler manifold $\widetilde{\mathcal{M}}$.
Thus the field strength of a holomorphic line bundle $\widetilde{L}$ is given by $\widetilde{\mathcal{F}}_{i\overline{j}} = \partial_{i} \overline{\partial}_{\bar{j}}
\widetilde{\phi}(z, \overline{z})$ where $\widetilde{\phi}(z, \overline{z})$ is a real smooth
function on $\mathbb{C}^3$. The identity \eq{dualdbi-id1} then demands to identify the real
function $\widetilde{\phi}(z, \overline{z})$ with the K\"ahler potential $\widetilde{K}(z, \overline{z})$
up to holomorphic gauge transformations
(see the footnote \ref{gluing}), i.e., \cite{hea,cy-hsy}
\begin{equation}\label{dualid-kaho}
    \widetilde{\phi}(z, \overline{z}) = \widetilde{K}(z, \overline{z}).
\end{equation}
Conversely, if the metric $\widetilde{\mathcal{G}}_{\mu\nu}$ is K\"ahler,
$\widetilde{\mathcal{F}}$ must be the curvature of a holomorphic line bundle.
As we pointed out in footnote \ref{i-sheaf}, it is necessary to replace holomorphic line bundles
with torsion free sheaves of rank one in order to include singular $U(1)$ gauge fields
such as $U(1)$ instantons. The torsion free sheaves fail to be a line bundle in real codimension
four \cite{ivno}, which are nothing but the ideal sheaves in our case.
The identity \eq{dualdbi-id2} similarly requires that NC $U(1)$ gauge fields should be a connection of
a NC holomorphic line bundle satisfying $\widehat{H}_{ij} = \widehat{H}_{\overline{i}\overline{j}} = 0$.
In particular, if the NC $U(1)$ gauge fields in Eq. \eq{dualdbi-id2} are NC Hermitian $U(1)$ instantons
obeying the Hermitian Yang-Mills equation
\begin{equation}\label{dual-nchym}
    \widehat{H}  = - * (\widehat{H} \wedge \widetilde{B}),
\end{equation}
the K\"ahler metric $\widetilde{\mathcal{G}}_{\alpha\overline{\beta}}$ in Eq. \eq{dualdbi-id1} describes
a CY manifold $\widetilde{\mathcal{M}}$ \cite{cy-hsy}.

The emergent CY manifold $\widetilde{\mathcal{M}}$ can be demonstrated on a more concrete basis.
As a counterpart of $\widehat{D}_a (y) = \mathcal{D}_{A} (p_a)$, let us introduce covariant NC momenta
defined by $\widehat{K}_a (u) \equiv \mathcal{D}_{\widetilde{A}} (\widetilde{p}_a)
= \widetilde{B}_{ab} \widehat{W}^b (u)$ where $\widetilde{p}_a = \widetilde{B}_{ab} u^b$.
Then they satisfy the commutation relation
\begin{equation}\label{ncc-kk}
    -i [\widehat{K}_a, \widehat{K}_b ]_{\widetilde{\star}} = -\widetilde{B}_{ab} + \widehat{H}_{ab}.
\end{equation}
To bear a close parallel to Eq. \eq{nc-vector}, let us consider the set of NC vector fields defined by
\begin{equation}\label{dualnc-vector}
\{ \widehat{Z}_a \equiv \mathrm{ad}_{\widehat{K}_a} \in \mathfrak{D}| \widehat{K}_a (u)
 = \widetilde{p}_a + \widehat{C}_a (u) \in \mathcal{A}_{\widetilde{\theta}}, \; a = 1, \cdots, 6 \}.
\end{equation}
One can apply the Lie algebra homomorphism \eq{lie-homo} to Eq. \eq{ncc-kk}
to yield the relation \cite{q-emg,hsy-review}
\begin{equation}\label{homo-dualh}
\mathrm{ad}_{\widehat{H}_{ab}}  =  [\widehat{Z}_a, \widehat{Z}_b] \in \mathfrak{D}.
\end{equation}
After all, the Hermitian Yang-Mills equation \eq{dual-nchym} can be transformed as \cite{cy-hsy,hsy-epjc09}
\begin{equation}\label{dve-zz}
    [\widehat{Z}_a, \widehat{Z}_b] = - \frac{1}{2} {T_{ab}}^{cd} [\widehat{Z}_c, \widehat{Z}_d],
\end{equation}
where ${T_{ab}}^{cd} = \frac{1}{2} {\varepsilon_{ab}}^{cdef} \widetilde{B}_{ef}$ and
$\widetilde{B} = \mathbf{1}_3 \otimes \sqrt{-1} \sigma^2$. Following the exactly same calculation
given in Ref. \cite{cy-hsy} (see Appendix B), one can show that the commutative limit
of Eq. \eq{dve-zz} is equivalent to geometric equations for spin connections given by
\begin{equation}\label{dual-spin}
    \widetilde{\omega}_{ab} = - \frac{1}{2} {T_{ab}}^{cd} \widetilde{\omega}_{cd}.
\end{equation}
Note that the spin connections $\widetilde{\omega}^a_{~\,b}$ are determined by solving the torsion-free
conditions:
\begin{equation}\label{dual-torsion}
    \widetilde{T}^a = d\widetilde{E}^a + \widetilde{\omega}^a_{~\,b} \wedge \widetilde{E}^b = 0
\end{equation}
for a six-dimensional manifold $\widetilde{\mathcal{M}}$ whose metric is given by
\begin{equation}\label{dual-6metric}
    ds^2 = \widetilde{\mathcal{G}}_{\mu\nu} (x) dx^\mu \otimes dx^\nu
    = \widetilde{E}^a \otimes \widetilde{E}^a.
\end{equation}
It is not difficult to show \cite{cy-hsy} that the six-dimensional manifold $\widetilde{\mathcal{M}}$
must be a CY manifold if its spin connections satisfy the relation \eq{dual-spin}.

\section{Mirror Symmetry of Emergent Geometry}

We showed that the doubling of symplectic structures due to the Hodge duality results in
two independent classes of NC $U(1)$ gauge fields by considering the Seiberg-Witten map \cite{ncft-sw}
for each symplectic structure. It may be emphasized that this result is a direct consequence
of the well-known Hodge duality stating the doubling of two-form vector spaces in Eq. \eq{double-2form}.
As a result, emergent gravity leads to an intriguing conclusion that the variety of six-dimensional
manifolds emergent from NC $U(1)$ gauge fields is doubled.
Note that a CY manifold $X$ always arises with a mirror pair $Y$ obeying
the mirror relation \cite{mirror-book}
\begin{equation}\label{mirror}
    h^{1,1} (X) = h^{2,1} (Y), \qquad h^{2,1} (X) = h^{1,1} (Y)
\end{equation}
where $h^{p,q} (M) = \dim H^{p,q} (M) \geq 0$ is a Hodge number of a CY manifold $M$.
When we conceive the emergent CY manifolds from the mirror symmetry perspective,
we cannot help investigating how the doubling for the variety of emergent geometry
is related to the mirror symmetry of CY manifolds.

Suppose that $M$ is a six-dimensional orientable manifold to equip a globally defined volume form.
This volume form allows us to define the Hodge dual operator $*: \Omega^k (M) \to \Omega^{6-k} (M)$
on a vector space
\begin{equation}\label{ext-vecsp}
\Lambda^* M = \bigoplus_{k=0}^6 \Omega^k (M),
\end{equation}
where $\Omega^k (M)$ is the space of $k$-forms on $T^* M$.
Consider a subspace of nondegenerate, closed two-forms and co-closed four-forms in $\Lambda^* M$
denoted by $S^2 (M)$ and $S^4 (M)$, respectively. Let us take a direct sum
\begin{equation}\label{direct-sum}
    S(M) \equiv S^2 (M) \oplus * S^4 (M).
\end{equation}
If $\omega \in S(M)$, then $\omega$ is a closed, $d\omega=0$, and nondegenerate two-form.
Therefore, $\omega$ is a symplectic structure on $M$.
According to the Hodge decomposition theorem \cite{besse}, any two-form $\omega \in \Lambda^2 M$
in Eq. \eq{double-2form} is decomposed as
\begin{equation}\label{ghodge-dec}
    \omega = \omega_H + d \alpha + \delta \beta,
\end{equation}
and thus the decomposition for a general symplectic two-form $\omega \in S(M)$ is given by
\begin{equation}\label{hodge-dec}
    \omega = \omega_H + d \alpha,
\end{equation}
where $\omega_H$ is a harmonic two-form and $\alpha \in \Omega^1 (M), \; \beta \in \Omega^3 (M)$.
A harmonic $k$-form $\omega_H \in \Omega^k (M)$ is defined by $\Delta \omega_H = 0$ where
the Laplace-Beltrami operator $\Delta: \Omega^k (M) \to \Omega^k (M)$ is defined by
\begin{equation}\label{lp-oper}
    \Delta = d\delta + \delta d.
\end{equation}
A $k$-form $\omega_H$ is harmonic if and only if $d\omega_H = 0$ and $\delta \omega_H = 0$.
Then $\omega_H$ is a unique harmonic representative in the $k$-th de Rham cohomology $H^k (M)$ \cite{besse}.
Note that the harmonic two-form $\omega_H = \omega_H^2 \oplus *\omega_H^4$ in Eq. \eq{hodge-dec} in general
consists of harmonic forms in $H^2 (M)$ and $H^4 (M)$. Similarly the one-form $\alpha \in \Omega^1 (M)$
in Eq. \eq{hodge-dec} contains $\alpha = - * \gamma$ with $\gamma \in \Omega^5 (M)$
as well as $\alpha = a$ in $\Omega^1 (M)$, that means $d\alpha = da \oplus * \delta \gamma$.
We remark that the Hodge decomposition on the exterior algebra \eq{ext-vecsp} is a canonical
decomposition given a globally defined volume form from which an oriented inner product
is defined. Hence it is necessary to consider the direct sum \eq{direct-sum} to realize
the Hodge decomposition \eq{hodge-dec} for a general symplectic structure since
$\omega_H^4 + \delta \gamma \in S^4 (M)$.

Since emergent gravity is based on the symplectic geometry or more generally a Poisson geometry \cite{q-emg},
it is necessary to exhaust, at least, all possible symplectic structures to realize
a complete emergent geometry. Therefore, it is demanded to consider the direct sum \eq{direct-sum}
to exhaust all possible symplectic structures. For instance, $\mathcal{F} = B + F$
in Eq. \eq{dec-ndfield} and $\widetilde{\mathcal{F}} = \widetilde{B} + \widetilde{F} = * (C + G)$
in Eq. \eq{dual-symp} belong to the vector space $S(M)$. In general, as we have shown before,
the vector space $\eq{direct-sum}$ can be understood as a deformation space of primitive symplectic
and dual symplectic structures $(B, \widetilde{B})$ which is locally described by a line bundle $L$
over $(M, B)$ and a dual line bundle $\widetilde{L}$ over $(M, \widetilde{B})$.
We verified how the doubling of symplectic structures in Eq. \eq{direct-sum} due to the Hodge duality
leads to two independent classes of NC $U(1)$ gauge fields and results in the doubling of
emergent geometry. For example, we showed in Sect. 3 that NC Hermitian $U(1)$ instantons
arise as a solution of the Hermitian Yang-Mills equation \eq{dual-nchym}
defined by dual NC $U(1)$ gauge fields $\widehat{C}_a (u)$ and give rise to CY manifolds
in the commutative limit, which are independent of CY manifolds emergent from the line bundle
$L$ over a symplectic manifold $(M, B)$. In other words, the variety of emergent CY manifolds
is doubled thanks to the Hodge duality $*: S^4 (M) \to S^{2} (M)$.

Note that the Euler characteristic of a CY manifold $M$ is given by \cite{mirror-book}
\begin{equation}\label{euler-cy}
\chi(M) = 2 \big(h^{1,1} (M) - h^{2,1} (M) \big).
\end{equation}
Since two classes of emergent CY manifolds are completely independent of each other,
it should be possible to arrange a pair of CY manifolds $(X, Y)$ such that $\chi (X) = - \chi(Y)$.
(A very similar doubling for the variety of CY manifolds was observed in \cite{cy-hym}.)
Because of the fact $h^{p, q} (M) = \dim H^{p,q} (M) \geq 0$, $\chi (X) = - \chi(Y)$ necessarily implies
the mirror relation \eq{mirror}. Consequently, the emergent gravity suggests a beautiful picture
that the mirror symmetry of CY manifolds simply originates from the doubling of
symplectic structures in Eq. \eq{direct-sum}. Furthermore, according to the Hodge decomposition theorem,
generic deformations of a symplectic structure can be written as the form \eq{hodge-dec},
in which $\omega_H = \omega_H^2 \oplus *\omega_H^4$ is a sum of harmonic forms in $H^2 (M)$ and $H^4 (M)$
and $d\alpha = da \oplus * \delta \gamma$ with $a \in \Omega^1 (M)$ and $\gamma \in \Omega^5 (M)$.

In summary, the generic deformation of a symplectic two-form can be written as the form
\begin{equation}\label{gen-sdef}
  \omega = (\omega_H^2 + da) + * (\omega_H^4 + \delta \gamma).
\end{equation}
Note that $\omega$ belongs to the vector space in Eq. \eq{direct-sum}, i.e., $\omega \in S^2 (M)
\oplus * S^4 (M)$ because $d\omega = d (\omega_H^2 + d a) + * \delta (\omega_H^4 + \delta \gamma) = 0$.
We showed that the deformations in $S^2 (M)$ are locally described by a line bundle $L \to M$ while
those in $* S^4 (M)$ are modeled by a dual line bundle $\widetilde{L}$ over $M$. Those two
deformations are independent of each other and result in two independent classes of CY manifolds.
Therefore we can derive two independent classes of CY manifolds from the deformations in Eq. \eq{gen-sdef}
and classify them according to their topological invariants. Since the Euler characteristic $\chi(M)$
of a CY 3-fold $M$ can have an arbitrary sign unlike the four-dimensional case in which
the Euler characteristic must be positive semi-definite, we may arrange a pair of CY manifolds $(X, Y)$
such that $\chi (X) = - \chi(Y)$ in which CY manifolds $X$ and $Y$ are emergent from the classes
$S^2 (M)$ and $S^4 (M)$, respectively. The formula \eq{euler-cy} indicates that the only solution
for $\chi (X) = - \chi(Y)$ is to satisfy the mirror relation \eq{mirror}. (Note that the Hodge diamond
for a CY 3-fold is determined by only two independent Hodge numbers $h^{1,1}$ and $h^{2,1}$
besides fixed ones $h^{3,0} = h^{0,3} = h^{0,0} = h^{3,3} = 1$.)
Therefore, the emergent gravity picture implies that the mirror symmetry of CY manifolds
can be understood as the Hodge theory for the deformations of symplectic and dual symplectic structures
characterized by Eq. \eq{gen-sdef}.

\section{Discussion}

The identification \eq{iden=fk} implies a general result \cite{hea,cy-hsy} that a holomorphic line bundle
with a nondegenerate curvature two-form is equivalent to a six-dimensional K\"ahler manifold.
A generalization to torsion free sheaves or ideal sheaves must be implemented to incorporate
singular $U(1)$ gauge fields such as $U(1)$ instantons.
Since the real function $\phi(z, \overline{z})$ will be determined by solving the equations of motion
of $U(1)$ gauge fields, it means that (generalized) Maxwell's equations for $U(1)$ gauge fields
on a holomorphic line bundle can be translated into Einstein's equations for a K\"ahler manifold.
A particular case was verified in \cite{hea,cy-hsy} that the Einstein equations
for CY $n$-folds for $n=2$ and $3$ are equivalent to the equations of motion for
the commutative limit of NC $U(1)$ instantons in four and six dimensions, respectively.
Recall that the metric for a K\"ahler manifold is basically determined by a single function, the so-called
K\"ahler potential, although the gluings described in the footnote \ref{gluing} must be implemented
to have a globally defined metric. As a result, the Ricci tensor of a K\"ahler manifold is extremely simple
and it is given by \cite{besse}
\begin{equation}\label{k-ricci}
    R_{\alpha\overline{\beta}} = - \frac{\partial^2 \ln \det \mathcal{G}_{\gamma\overline{\delta}}}
    {\partial z^\alpha \partial \overline{z}^{\bar{\beta}}}.
\end{equation}
Using the identity \eq{iden=fk}, it must be possible to relate the Ricci tensor \eq{k-ricci} to
some equations of $U(1)$ gauge fields on a holomorphic line bundle. It will be interesting to
find an explicit form of the equations for holomorphic $U(1)$ gauge fields.

So far we have assumed that a complex manifold $M$ is noncompact, e.g., $\mathbb{C}^3$.
It is desirable to generalize the results in this paper to compact complex manifolds, e.g.,
$T^6, \, T^2 \times K3$, and $\mathbb{CP}^3$, which are all compact symplectic
(i.e., K\"ahler) manifolds. We can put a holomorphic line bundle on such a compact K\"ahler manifold.
Then, similarly to the noncompact case, the line bundle will deform an underlying symplectic
(i.e., K\"ahler) structure of the base manifold and end in a dynamical symplectic manifold.
The resulting symplectic structure can be identified with the K\"ahler form of a K\"ahler
manifold emergent from the holomorphic line bundle over a compact complex manifold.
Therefore, we still have the local identification \eq{iden=fk} even for a compact manifold.
However, an explicit construction of Poisson algebras and covariant connections on a compact
K\"ahler manifold will be much more difficult than a noncompact case.
In particular, the gluing of coordinate patches for a holomorphic atlas of a compact manifold,
described in the footnote \ref{gluing}, will be more nontrivial compared to, e.g., $\mathbb{C}^3$.
The quantization of a compact K\"ahler manifold will also be a more challenging issue.
Thus a sophisticated mathematical tool for emergent geometry would be requested
for the compact case. Nevertheless, the conclusion for the noncompact case will be true even
for compact K\"ahler manifolds because main features such as Eqs. \eq{direct-sum} and \eq{hodge-dec}
invariably hold for any symplectic manifold.

In four dimensions, it has been possible to accomplish an explicit test of emergent gravity with known
solutions in gravity and gauge theory \cite{hsy-inst,lrya13}.
In higher dimensions, it becomes more difficult to obtain
an explicit solution in gravity as well as gauge theory. Fortunately, some solutions
in six dimensions are explicitly known for Ricci-flat K\"ahler manifolds \cite{exsol-cy6} and
NC Hermitian $U(1)$ instantons \cite{bkp-hbps,nc-inst6}. Therefore, it will be interesting to examine
an explicit test of six-dimensional emergent gravity for the known solutions
in both gravity and gauge theory.

\section*{Acknowledgments}

This work was supported by the National Research Foundation of Korea (NRF) grant funded
by the Korea government (MOE) (No. NRF-2015R1D1A1A01059710).

\end{document}